\documentclass[]{amsart}
\usepackage[lmargin=3cm,tmargin=3.5cm,bmargin=3.5cm,rmargin=3cm]{geometry} 
\usepackage[table]{xcolor}
\usepackage{tabularx}
\usepackage{mathrsfs,amsmath, amsfonts, amssymb, amscd}
\usepackage{amsaddr}
\usepackage{graphicx,graphics,epsfig}
\usepackage{subfig}
\usepackage{amssymb}
\usepackage{amsfonts}
\usepackage{amsmath}
\usepackage{epsfig}
\usepackage{float}
\usepackage{setspace}
\usepackage{tikz}
\usepackage{enumerate}
\usepackage[colorlinks=true,urlcolor=blue, citecolor=red,linkcolor=blue,linktocpage,pdfpagelabels, bookmarksnumbered,bookmarksopen]{hyperref}
\usepackage{marginnote}
\usepackage{epstopdf}

\newtheorem{theorem}{Theorem}[section]

\newtheorem{corollary}[theorem]{Corollary}
\newtheorem{lemma}[theorem]{Lemma}
\newtheorem{remark}[theorem]{Remark}

\newcommand{\p}{\partial}

\newcommand{\R}{{\mathbb R}}

\newcommand{\C}{{\mathbb C}}
\newcommand{\T}{{\mathbb T}}

\newcommand{\dem}[1]{{\it \noindent \textbf{Proof}.} #1\hfill $\blacksquare$\smallskip}

\title[]{Asymptotic behavior of the Schr\"odinger-Debye system with refractive index of square wave amplitude}

\author[A. J. Corcho]{\bf Ad\'an J. Corcho}
\address{Instituto de Matem\'atica, Universidade Federal do Rio de Janeiro.\\ 
Centro de Tecnologia - Bloco C. Cidade Universit\'aria.\\
Ilha do Fund\~ao  21941-909.  Rio de Janeiro - RJ, Brazil.}
\email{adan@im.ufrj.br}

\author[J. C. Cordero]{\bf Juan C.  Cordero}
\address{Departamento de Matem\'aticas y Estad\'istica.\\ 
Universidad Nacional de Colombia-Sede Manizales.\\
Manizales-Colombia.}
\email{jccorderoc@unal.edu.co}

\thanks{A. J. Corcho was partially supported by CAPES and CNPq (through 309752/2013-2), Brazil}
\thanks{J. C. Cordero was partially supported by the CONVOCATORIA PARA LA MOVILIDAD INTERNACIONAL DE LA UNIVERSIDAD NACIONAL DE COLOMBIA 2016-2018 (movilidad  6685)}


\date{\today}

\begin{document}

\setcounter{page}{1}

\begin{abstract}
We obtain local well-posedness for the one-dimensional Schr\"odinger-Debye interactions in nonlinear optics in the spaces $L^2\times L^p,\; 1\le p < \infty$. When $p=1$ we show that the local solutions extend globally. In the focusing regime, we consider a family of solutions $\{(u_{\tau}, v_{\tau})\}_{\tau>0}$ in $ H^1\times H^1$ associated to an initial data family $\{(u_{\tau_0},v_{\tau_0})\}_{\tau>0}$ uniformly bounded in $H^1\times L^2$, where $\tau$  is a small response time parameter. We prove  prove that $(u_{\tau}, v_{\tau})$ converges to $(u, -|u|^2)$ in the space $L^{\infty}_{[0, T]}L^2_x\times L^1_{[0, T]}L^2_x$ whenever $u_{\tau_0}$ converges to $u_0$ in $H^1$ as long as $\tau$ tends to 0, where $u$ is the solution of the one-dimensional cubic non-linear Schr\"odinger equation with initial data $u_0$. The convergence of $v_{\tau}$ for $-|u|^2$ in the space $L^{\infty}_{[0, T]}L^2_x$ is  shown under compatibility conditions of the initial data. For non compatible data we prove convergence except for a corrector term which looks like an initial layer phenomenon.  
\end{abstract}

\maketitle
\section{\bf{Introduction}}
We consider the family of systems, labelled by a parameter $\tau >0$, given by the following coupling equations: 
\begin{equation}\label{SD}
\begin{cases}
i\partial_tu+ \frac12\partial^2_xu=uv, & (x,t)\in \mathbb{R}\times \mathbb{R}^+,\\
\tau\partial_t v + v = \lambda |u|^{2},& \qquad\; 0 < \tau \ll 1, \\
u(x,0)=u_{0}(x),\quad v(x,0)=v_{0}(x).&
\end{cases}
\end{equation}
This model describes the propagation of an electromagnetic wave through a non-resonant medium, whose nonlinear polarization lags behind the induced electric field. The complex function $u$ 
represents  the electromagnetic wave, the real function $v$ measures the medium refraction index while $\tau$ is a small response time parameter of the variation of $v$ due to the electromagnetic field. For  more   physical details we refer the book \cite{Newell}. This system of partial differential equations is known in the literature as  the Schr\"o\-dinger-Debye (SD) system. It also has physical interest for data defined in the euclidean spaces $\R^2$ or $\R^3$. 

We refer to the system \eqref{SD} as focussing or defocussing according to the value of $\lambda$ being -1 or 1, respectively. This is in accordance with the nonlinear Schrodinger equation. 

The only conservation law known for SD system is the $L^2$-norm of the solution $u$, that is,
\begin{equation}\label{Conservation-Law-L2}
\int_{-\infty}^{+\infty}|u(x,t)|^{2}dx=\int_{-\infty}^{+\infty}|u_{0}(x)|^{2}dx.
\end{equation}

We realize that the  system  (\ref{SD}) is formally reduced (as $\tau\to0$) to the cubic non-linear Schr\"odinger equation (cubic-NLS):
\begin{equation}\label{CNLS}
\begin{cases}
i\partial_tu+ \tfrac{1}{2}\partial^2_x u = \lambda |u|^2u,\quad (x,t)\in \R\times \mathbb{R}^+,\medskip \\
u(x,0)=u_0(x),
\end{cases}
\end{equation}
which would be representing an instantaneous polarization response in the absence of delay  ($\tau=0$).

The system (\ref{SD}) can be decoupled, by solving the second equation with respect to $v$,
\begin{equation}\label{equation-v}
v(\cdot, t)= e^{-t/\tau}v_0(x)+ \tfrac{\lambda}{\tau}\int_0^t\,e^{-(t-s)/\tau}|u(\cdot, s)|^2\,ds,
\end{equation}
to obtain the following integro-differential equation for $u$:
\begin{equation}\label{SD-IDF}
\begin{cases}
i\partial_tu+\tfrac{1}{2}\partial_x^2 u = e^{-t/\tau}uv_0(x)
+\frac{\lambda}{\tau}u\displaystyle \int_0^t\,e^{-(t-s)/\tau}|u(\cdot, s)|^2ds,&  x\in \R,\; t\ge 0,\\
u(x,0) = u_0(x).
\end{cases}
\end{equation}

We note that for data $v_0\in L^1(\R)$, the integral expression \eqref{equation-v} suggests that the solution $v(\cdot, t)$ remains in  $L^1(\R)$ whenever the solution 
$u(\cdot, t)\in L^2(\R)$. So, it is natural to ask if it is possible to develop a local theory for the system SD in the space $L^2\times L^1$. In fact, the first goal in this work
is to establish local well-posedness results for data in the space  $L^2\times L^p,\; 1\le p< \infty$, which can be extended globally in the case $p=1$. 

Another system which is somewhat more general than (\ref{SD}) is named Maxwel-Debye system:

\begin{equation}\label{MD}
\begin{cases}
i\partial_tu+ \frac12(\partial^2_x+\partial^2_y)u=uv, \\
\tau(\partial_t v+\partial_{\xi}v) + v = \lambda |u|^{2},\\
u(x,0)=u_{0}(x),\quad v(x,0)=v_{0}(x),&
\end{cases}
\end{equation}
which is in a reference frame moving at the velocity $c$, with $\xi=ct-z$. Using the corresponding formulation like \eqref{SD-IDF}, Bid\'egaray proved in \cite{Bidegaray1} that for a small enough time $T$, the solutions $u_{\tau}$ of the Maxwel-Debye system converge strongly (in the topology of the space $C\big([0, T];\, H^s_x\big)$) to the corresponding solution $u$ of the equation \eqref{CNLS}, whenever the initial data $(u_0, v_0)$ belong to the Sobolev space $H^s(\R) \times H^s(\R)$ and $s>5/2$. For regularity $s=1$ she could also show convergence but in a weaker sense. One shall notice that these results nothing says about the convergence of refraction solutions $v_{\tau}$. 

Because of \eqref{Conservation-Law-L2}, our second goal in this work is to study the asymptotic behaviour of the  solutions of system SD in the topology of $C\big([0,T];\,L^2(\mathbb{R})\big)$ for the component  $u_{\tau}$ and with an appropriated topology for the refraction solutions  $v_{\tau}$. Of course, if we expect a strong convergence result in $C([0,T];L^2(\mathbb{R}))$ for solutions  $v_{\tau}$ it is natural to impose a compatibility condition on the initial data, like
\begin{equation}
v_0=\lambda\vert u_0 \vert^2\quad \text{or} \quad \lim\limits_{\tau\to 0}\big\|{v_0}_{\tau}-\lambda\vert {u_0}_{\tau} \vert^2\big\|_{L^2}=0\quad \text{if the data vary with}\; \tau.
\end{equation}

For instance, if  $u_{0\tau}\equiv0$ then 
\begin{equation}
u_{\tau}(x, t)\equiv0\quad\text{and}\quad v_{\tau}(x,t)=e^{-t/{\tau}}v_{0\tau},
\end{equation}
so  
\begin{equation}
\Vert v_{\tau}(x,t) \Vert_{L_t^{\infty}L^2_x}=\Vert v_{0\tau}\Vert_{L_x^2}.
\end{equation}
Hence,  we do not have much chance of show convergence convergence in the space $L_T^{\infty}L^2_x\times L_T^{\infty}L^2_x$ without assu\-ming that
$\Vert v_{0\tau}\Vert_{L^2_x} \to 0$ as $\tau\to0$. For non compatible initial data we will see that an initial layer phenomenon appears; however, in this situation, the best result obtained in this work is the convergence of focusing ($\lambda=-1$) solutions $v_{\tau}$ for $|u|^2$ in the topology of the mixed space $L^1\big([0,T];\,L^2(\mathbb{R})\big)$. Another important point is that our converge results are valid for any fixed time interval $[0, T]$ and they improve considerably the previous ones obtained in \cite{Bidegaray1} in the one-dimensional case.

Before to state the main results we will briefly review previous results regarding well-posedness for the Schr\"odinger-Debye system. 

\subsection{Overview about well-posednes}
The first results concerning well-posedness for the Schr\"odinger-Debye system \eqref{SD} with data defined in Sobolev spaces can be seen in \cite{Bidegaray1,Bidegaray2,Corcho-Linares}. The more general results known to date for data defined in Sobolev spaces on the line were given in
\cite{Corcho-Matheus}, which reads as follows:

\medskip 
{\noindent \bf Theorem A.} 
{\it For any $(u_0,v_0)\in H^s(\R) \times H^{\kappa}(\R)$, with $s$ and $\kappa$ verifying the conditions:
$$|s|-\frac12\le \kappa < \min\left\{s+\frac12,\; 2s+\frac12\right\}\; \text{and}\; s> -\frac14,$$
there exists a time $T=T(\|u_0\|_{H^{s}}, \|v_0\|_{H^{\kappa}})>0$ and a unique solution $(u(t),v(t))$ of the initial value problem
(\ref{SD}) in the time interval $[0,T]$, satisfying
$$(u, v)\in C\left([0,T]; H^s(\R) \times H^{\kappa}(\R)\right).$$
Moreover, the map $(u_0,v_0) \longmapsto (u(t),v(t))$ is locally Lipschitz. In addition, when $-3/14< s=\kappa \le 0$, the
local solutions can be extended to any time interval $[0,T]$.}

\begin{figure}[htp]
\centering
\begin{tikzpicture}
\draw[very thin,color=gray, dashed] (-1.9,-1.9) grid (3.9,3.9);
\draw[->] (-2,0)--(4,0) node[below] {$s$};
\draw[->] (0,-1.5)--(0,4) node[right] {$\kappa$};
\filldraw[color=gray!40]
(-0.25,-0.25)--(-0.25,0)--(0,0.5)--(3.5,4)--(4,4)--(4,3.5)--(0,-0.5)--(-0.25,-0.25);
\draw[thick, dashed](-0.25,-0.25)--(-0.25,0)--(0,0.5)--(3.5,4);
\draw[thick](4,3.5)--(0,-0.5)--(-0.25,-0.25);
\draw[very thin](-0.25,0)--(0.5,0);
\draw[very thin] (0,-0.5)--(0,0.5);
\node at (2,2){\small{$\boldsymbol{\mathcal{W}_A}$}};
\node at (-0.9,0.4){\small{$\kappa=2s+\frac12$}};
\node at (1.2,2.7){\small{$\kappa=s+\frac12$}};
\node at (1.2,-0.4){\small{$\kappa=|s|-\frac12$}};
\end{tikzpicture}
\caption{l.w.p. regularity given in Theorem A}\label{Region-A}
\end{figure}
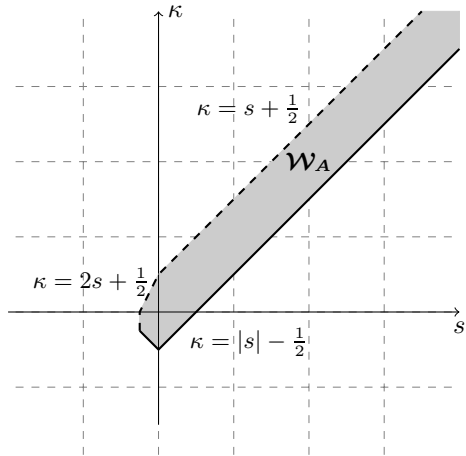

These results were obtained by applying a fixed-point procedure to the Duhamel formulation asso\-ciated to the system  \eqref{SD} and  using the structure of  Bourgain's spaces  associated to  the system. The global results are based on a good control of the $L^2$-norm of the solution $v$, which provides global well-posedness in $L^2\times L^2$. Global well-posedness below $L^2$-regularity is obtained via the \emph{I-method} introduced by Colliander, Keel, Staffilani, Takaoka and Tao in \cite{Imethod}. As pointed out in concluding remark 4.2 of \cite{Corcho-Oliveira-Silva}, global well-posedness in $H^1(\R)\times H^1(\R)$ regularity, included in Theorem A, can be obtained in this space. 

Figure \ref{Region-A} represents the region $\mathcal{W}_A$ in the $(s, \kappa)$ plane,  corresponding to the sets of Sobolev indices for which local well-posedness (l.w.p.) has been established, in \cite{Corcho-Matheus}, as described in Theorem A.

On the other hand, the results in  \cite{Angulo-Corcho-Hakkaev, Corcho-Matheus} ensure the following analogous local well-posed theory for periodic initial data. 

\medskip 
{\noindent \bf Theorem B.} 
{\it For any $(u_0,v_0)\in H^s(\T) \times H^{\kappa}(\T)$, with $s$ and $\kappa$ verifying the conditions:
	\begin{equation}\label{local-theorem-periodic-a}
	0\le \kappa \le 2s\;\;\; and \;\;\; |\kappa-s|\le 1,
	\end{equation}
	there exist a positive time $T=T(\|u_0\|_{H^s_{per}}, \|v_0\|_{H^{\kappa}_{per}})$ and a unique solution $(u(t),v(t))$ of the initial value problem
	(\ref{SD}) in the time interval $[0,T]$, satisfying, 
	$$(u, v)\in C\left([0,T]; H^s(\T) \times H^{\kappa}(\T)\right).$$
	Moreover, the map $(u_0,v_0) \longmapsto (u(t),v(t))$ is locally Lipschitz.}

\begin{figure}[htp]
	\centering
	\begin{tikzpicture}
	\draw[very thin,color=gray, dashed] (-1,-1) grid (3.8,4.8);
	\draw[->] (-1,0)--(4,0) node[below] {$s$};
	\draw[->] (0,-0.5)--(0,5) node[right] {$\kappa$};
	\filldraw[color=gray!40](0,0)--(1,2)--(4,5)--(4,3)--(1,0)--(0,0);
	\draw[thick](4,5)--(1,2)--(0,0)--(1,0)--(4,3);
	\node at (2.3,2.3){\small{$\boldsymbol{\mathcal{W}_B}$}};
	\node at (0.05,1.3){\small{$\kappa=2s$}};
	\node at (1.25,3.2){\small{$\kappa=s+1$}};
	\node at (3.15,1.2){\small{$\kappa=s-1$}};
	\end{tikzpicture}
	\caption{l.w.p. regularity given in Theorem B}\label{Region-B}
\end{figure}
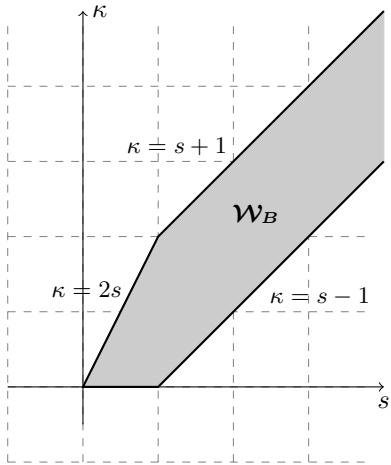

\subsection{Main results}
We establish now the results obtained in our work for the Cauchy problem \eqref{SD} with initial data defined on the line. The first result establishes well-posedness for data in $L^2\times L^p,\; 1\le p <\infty$ and the last two theorems are concerning the convergence of solutions of system SD to the cubic-NLS equation in different topologies. 

\medskip
\begin{theorem}[\textbf{Local well-posedness in $\boldsymbol{L^2\times L^p}$}]\label{LWP-SD} Consider the system (\ref{SD}) with $\lambda =\pm 1$ and initial data $(u_0,v_0)\in L^2(\R)\times L^p(\R)$ for $1\le p < \infty$. Then, there exists a positive time $T=T(\tau,\, \|u_0\|_{L^2},\, \|v_0\|_{L^p})$ and a unique solution 
$(u(\cdot, t),\, v(\cdot, t))\in C\left([0,T];\, L^2\times L^p\right)$ satisfying
\begin{equation}\label{Th-local-a}
u\in L^{\frac{4p}{2-p}}_TL^{\frac{p}{p-1}}_x \cap  L^{\frac{4p}{p-1}}_TL^{2p}_x \;\; \text{if}\;\; 1\le p < 2,
\end{equation}
\begin{equation}\label{Th-local-b}
u\in L^{2p}_TL^{\frac{2p}{p-2}}_x\cap L^{\frac{4p}{p-1}}_TL^{2p}_x\;\; \text{if}\;\; 2 \le p < \infty . 
\end{equation}
Moreover, for all $0< T'<T$, there exists a neighborhood $U'\times V'$ of $(u_0, v_0)$ in $L^2\times L^p$ such that the map 
$(u_0, v_0) \longmapsto (u(\cdot, t),\, v(\cdot, t))$ from $U'\times V'$ into the class defined by 
\eqref{Th-local-a} - \eqref{Th-local-b} with $T'$ instead of $T$ is Lipschitz. 
\end{theorem}

\medskip
\begin{corollary}\label{GWP-SD}
For any $(u_0, v_0) \in L^2\times L^1$ the local solution given by Theorem \ref{LWP-SD} can be extended to any time interval $[0, T]$.
\end{corollary}

\medskip
\begin{theorem}[$\boldsymbol{L^\infty_TL^2_x\times L^1_TL^2_x}$ \textbf{- convergence}]\label{Convergence-L2-A} 
Consider the system \eqref{SD} with $\lambda =-1$ and a family of initial data $\big\{({u_0}_{\tau}, {v_0}_{\tau})\big\}_{0< \tau <1}$ in the space $H^1\times H^1$ such that 
\begin{enumerate}
\item[(i)] $\lim\limits_{\tau\to0}\Vert{u_0}_{\tau}-u_0\Vert_{H^1}=0$\quad for some $u_0\in H^1$.
\end{enumerate}
Suppose further that 
\begin{enumerate}
\item[(ii)] $\displaystyle \sup_{0<\tau <1}\|{v_0}_{\tau}\|_{L^2} < \infty $\quad\text{or}\\
 	
\item[(iii)]$\displaystyle \|{v_0}_{\tau}\|_{L^2}=O(\tau^{-1+\gamma})$, with $0<\gamma \ll 1$,\quad\text{if}\quad $\displaystyle \sup_{0<\tau <1}E(0)\le 0$.
\end{enumerate}
If $(u_{\tau}, v_{\tau})$ denotes the corresponding global solution in $H^1\times H^1$ for the system \eqref{SD} with initial data $({u_0}_{\tau}, {v_0}_{\tau})$ and $u$ is the corresponding global solution in $L^2$ for cubic-NLS equation \eqref{CNLS} with initial data $u_0$, then for all $T^*>0$ we have
\begin{equation}
\lim\limits_{\tau\to 0}\|u_{\tau}-u\|_{L^{\infty}_{T^*}L^2_x}=0
\end{equation}
and
\begin{equation}
\lim\limits_{\tau\to 0}\big\|v_{\tau}+|u|^2\big\|_{L^1_{T^*}L^2_x}=0.
\end{equation}
\end{theorem}

We note that, in the case of hypothesis (iii) of Theorem \ref{Convergence-L2-A}, the norms of initial data ${v_0}_{\tau}$ can grow infinitely with a rate of order $1/\tau^{1-}$ when $\tau$ is going to zero, however the convergence in $L^\infty_{T^*}L^2_x\times L^1_{T^*}L^2_x$ is ensured. We will see that the proof in this case is essentially the same as that given under hypothesis (ii).

The next result is concerning to the convergence in $L^\infty_{T^*}L^2_x$ for the solution $v_{\tau}$ of the system.

 \begin{theorem}[$\boldsymbol{L^\infty_TL^2_x\times L^\infty_TL^2_x}$ \textbf{- convergence}]\label{Convergence-L2-B} 
 	With the same hypotheses of Theorem \ref{Convergence-L2-A}, then for all $T^*>0$ we have
 	\begin{equation}\label{Convergence-L2-B-1}
 	\lim\limits_{\tau\to 0}\|u_{\tau}-u\|_{L^{\infty}_{T^*}L^2_x}=0
 	\end{equation}
 	and
 	\begin{equation}\label{Convergence-L2-B-2}
 	\lim\limits_{\tau \to 0}\big\| v_{\tau} +|u|^2-e^{-t/\tau}({v_0}_{\tau} + |u_0|^2)\big\|_{L^{\infty}_{T^*}L^2_x}=0.
 	\end{equation}
 	Moreover,
	\begin{equation}\label{Convergence-L2-B-3}
 	\lim\limits_{\tau\to 0}\big\|v_{\tau}+|u|^2\big\|_{L^\infty_{T^*}L^2_x}=0
 	\end{equation}
whenever
\begin{equation}
\lim\limits_{\tau\to 0}\big\Vert v_{0\tau}+\vert  u_0\vert^2\big\Vert_{L^2}=0.
 \end{equation}
 \end{theorem}

\begin{remark}
Note that $\omega_{\tau}(x,t):=-e^{-t/\tau}({v_0}_{\tau} + |u_0|^2)$ in \eqref{Convergence-L2-B-2} is the solution of the homogeneous Cauchy problem:
$$
\begin{cases}
\tau\partial_t\omega(x,t) +\omega(x,t) =0,\\
\omega(x,0)={v_0}_{\tau} + |u_0|^2.
\end{cases}
$$
The term $\omega_{\tau}$ is a fitting corrector which appears as an initial layer phenomenon because of non-compatibility conditions, similar to behavior of the Zakharov system when it captures the dynamics of the cubic non-linear Schr\"odinger equation in the adiabatic limit (cf. \cite{Added-Added} ).  
\end{remark}

The paper is organized as follows: in  Section $2$ we recall some notation and present some preliminary and necessary results. Section $3$ is devoted to the proof of local theory in the spaces $L^2\times L^p,\; 1\le p < \infty$  and finally, in Section $4$, we prove the convergence results.     

\section{{\bf Preliminaries}}
In this section we present the key background for the development of the proofs of our results. 
\subsection{Integral formulation, the pseudo Hamiltonian and the global smooth effects}
We will denote by
\begin{equation}
S(t)=e^{it\partial_x^2/2}, \quad D_{\tau}(t)=e^{-t/\tau}
\end{equation}
the groups associated to the Schr\"odinger and Decay equations respectively. Then the solution for the system SD obey the following integral equations:
\begin{equation}
\begin{cases}
\displaystyle u_{\tau}(x,t)=S(t)u_{0\tau}-i\int_0^{t}S(t-s)u_{\tau}(x,s)v_{\tau}(x,s)ds,\medskip\\
\displaystyle v_{\tau}(x,t)=D_{\tau}(t)v_{0\tau}+\tfrac{\lambda}{\tau}\int_0^{t}D_{\tau}(t-s)\vert u_{\tau}(x,s)\vert^2ds.
\end{cases}
\end{equation}
Here, $({u_0}_{\tau}, {v_0}_{\tau})$ are the initial data, and we have written $(u_{\tau},v_{\tau})$ for solutions of the SD system in order to consider the variation of them with respect to the delay parameter  $\tau$.

Similarly, the solution $u$ for the cubic NLS equation with initial data $u_0$ verifies
\begin{equation}
u(x,t)=S(t)u_0-i\int_0^{t}S(t-s)\lambda\vert u(x,s)\vert^2 u(x,s)ds.
\end{equation}
 
As we have already seen, the flow of the system \eqref{SD} preserves the $L^2$-norm of the solution $u_{\tau}$, that is,
\begin{equation}\label{Conservation Law}
\int_{-\infty}^{+\infty}|u_{\tau}(x,t)|^{2}dx=\int_{-\infty}^{+\infty}|u_{0\tau}(x)|^{2}dx.
\end{equation}

Also, the following pseudo-Hamiltonian structure holds:
\begin{equation}\label{Energy-1}
\frac{d}{dt}E_{\tau}(t)=2\lambda\tau\int_{-\infty}^{+\infty}(\partial_tv_{\tau}(x,t))^{2}dx,
\end{equation}
where
\begin{equation}\label{Energy-2}
E_{\tau}(t)=\int_{-\infty}^{+\infty}\Big(|\partial_xu_{\tau}|^{2}+\lambda |u_{\tau}|^{4}-\lambda \tau^{2}(\partial_tv_{\tau})^2\Big)dx
=\int_{-\infty}^{+\infty}\Big(|\partial_xu_{\tau}|^{2}+2v_{\tau}|u_{\tau}|^2 -\lambda v_{\tau}^2\Big)dx.
\end{equation}

$E_{\tau}$ is not conserved, however we can immediately infer its monotonicity, which depend on the sign of $\lambda$: increases in time when $\lambda=1$ or decreases  when $\lambda=-1$.

The energy integral \eqref{Energy-2}  is well defined as long as  $u_{\tau} \in H^1(\R)$ and $v_{\tau} \in L^2(\R)$, but unfortunate 
this regularity is not covered by the local theory developed in Theorem A. 

For $T>0$ a fixed time, we are going to write
\begin{equation}
 \Vert f\Vert_{L^1_T}:=\Vert f\Vert_{L^1_{[0,T]}}=\int_0^T\vert f(t)\vert dt,\quad\Vert f\Vert_{L^\infty_T}:=\sup_{t\in[0,T]}\vert f(t)\vert, \quad \Vert f\Vert_{L^\infty_t}:=\sup_{t\ge0} \vert f(t)\vert
\end{equation}
and the symbol $\Vert \cdot \Vert_{L_t^pL_x^q}$ will indicate the typical norm of a mixed space.

In particular, note that
\begin{equation}
\Big\Vert \tfrac{1}{\tau}D_{\tau}(t-\cdot_{s} ) \Big\Vert_{L^1_{[0,t]}}=\frac{1}{\tau}\int_0^{t}D_{\tau}(t-s)ds=1-D_{\tau}(t)<1
\end{equation}
for all $\tau>0$ and $t\ge 0$. This property of $D_{\tau}$ will be used when we have to deal with a control of the square wave amplitudes in the convergence results

Now we recall  the Strichartz estimate for the free Schr\"odinger group $\displaystyle S(t)$ in the euclidean space $\R$.

\begin{lemma}[\textbf{Strichartz estimates} \cite{Cazenave-Book}] Let $(p_1,q_1)$ and $(p_2,q_2)$ be two pairs of admissible exponents for $S(t)$ in $\R$; that is, both satisfying
	the condition
	\begin{equation}\label{Strichart-Admissible}
	\frac{2}{p_i}=\frac12 -\frac 1{q_i}\quad \text{and} \quad 2\le q_i\le \infty\quad (i=1,2).
	\end{equation}
	Then, for any $0<T\leq \infty$, we have
	\begin{equation}\label{Strichart-homogeneous}
	\|S(t)f\|_{L^{p_1}_TL^{q_1}_x}\le c\|f\|_{L^2(\R)},
	\end{equation}
	as well as the non-homogeneous version
	\begin{equation}\label{Strichart-non-homogeneous}
	\left\|\int_0^tS(t-s)g(\cdot, s)ds\right\|_{L^{p_1}_TL_x^{q_1}}\le c \|g\|_{L^{p'_2}_TL_x^{q'_2}},
	\end{equation}
	where $1/p_2+1/p_2'=1$, $1/q_2 + 1/q'_2=1$. The constants in both inequalities are independent of $T$. 
\end{lemma}
 
\subsection{A priori estimates for solutions in the space $\boldsymbol{H^1\times L^2}$.}
In this section we consider the system \eqref{SD} in focusing regime ($\lambda =-1$). The next results describe how the global solutions of \eqref{SD} in $H^1\times H^1$ grow in the space $H^1\times L^2$ with respect to the parameter $\tau$, when extra hypotheses are put on the initial data. Remember that no local theory is known for system \eqref{SD} in the space $H^1\times L^2$ in the one-dimensional case for spatial dimension, so we need to take data in $H^1\times H^1$. 

\begin{lemma}\label{apriori-estimate-H1L2}
If $\big\{({u_0}_{\tau}, {v_0}_{\tau})\big\}_{0< \tau <1}$ is a family of data in the space $H^1\times H^1$  such that
\begin{equation}
\sup_{0< \tau < 1}(\|{u_0}_{\tau}\|_{H^1}+\|{v_0}_{\tau}\|_{L^2})<\infty,
\end{equation}
then
\begin{equation}
\sup_{0< \tau < 1}(\|{u_{\tau}}\|_{L^{\infty}_tH^1_x}+\|v_{\tau}\|_{L^{\infty}_tL^2_x})<\infty. 
\end{equation}
\end{lemma}

\dem{Let $r:=\displaystyle \sup_{0< \tau < 1}(\|{u_0}_{\tau}\|_{H^1}+\|{v_0}_{\tau}\|_{L^2})$. As $E_{\tau}$ is decreasing we have
$$\|\p_x u_{\tau}(\cdot, t)\|_{L^2}^2+ \|v_{\tau}(\cdot, t)\|_{L^2}^2\le -2\int_{-\infty}^{+\infty}v_{\tau}(x, t)|u_{\tau}(x, t)|^2dx + E_{\tau}(0)\quad\forall\,t\ge0.$$
Using H\"older's and Young inequality, the Gagliardo-Nirenberg interpolation inequality and \eqref{Conservation Law} we obtain
  \begin{equation*}\label{apriori-estimate-H1L2-proof-a}
  \begin{split}
  \|\p_x u_{\tau}(\cdot, t)\|_{L^2}^2+ \|v_{\tau}(\cdot, t)\|_{L^2}^2&\le 2\|v_{\tau}(\cdot, t)\|_{L^2}\|u_{\tau}(\cdot, t)\|_{L^4}^2 + E_{\tau}(0)\\
  &\le \frac{1}{2}\|v_{\tau}(\cdot, t)\|^2_{L^2} + 2\|u_{\tau}(\cdot, t)\|_{L^4}^4 + E_{\tau}(0)\\
  &\le \frac{1}{2}\|v_{\tau}(\cdot, t)\|^2_{L^2}  + 2C_{gn}^4\|{u_0}_{\tau}\|_{L^2}^3\|\p_x u_{\tau}(\cdot, t)\|_{L^2} + E_{\tau}(0)\\
  &\le \frac{1}{2}\left( \|\p_x u_{\tau}(\cdot, t)\|_{L^2}^2 +\|v_{\tau}(\cdot, t)\|^2_{L^2} \right) +  2C^8_{gn}\|{u_0}_{\tau}\|_{L^2}^6+ E_{\tau}(0).
  \end{split}
  \end{equation*}
Hence, for all $t\ge 0$, we conclude that 
	\begin{equation}\label{apriori-estimate-H1L2-proof-b}
	\begin{split}
	\|\p_x u_{\tau}(\cdot, t)\|_{L^2}^2+ \|v_{\tau}(\cdot, t)\|_{L^2}^2&\le 4C^8_{gn}\|{u_0}_{\tau}\|_{L^2}^6+ 2E_{\tau}(0)\\
	&\le  4C^8_{gn}r^6+ 2E_{\tau}(0).
	\end{split}
	\end{equation}
	On the other hand, similar estimates give
	\begin{equation}\label{apriori-estimate-H1L2-proof-c}
	\begin{split}
	E_{\tau}(0)&\le \|\p_x {u_0}_{\tau}\|_{L^2}^2 +2\|{v_0}_{\tau}\|_{L^2}\|{u_0}_{\tau}\|^2_{L^4}+\|{v_0}_{\tau}\|^2_{L^2}\\
	&\le \|\p_x {u_0}_{\tau}\|_{L^2}^2 +2C_{gn}^2\|{v_0}_{\tau}\|_{L^2}\|{u_0}_{\tau}\|_{L^2}^{3/2}\|\p_x{u_0}_{\tau}\|^{1/2}_{L^2}+\|{v_0}_{\tau}\|^2_{L^2}\\
	&\le r^2 + 2C_{gn}^2r^3 +r^2.
	\end{split}
	\end{equation}
Then, combining the conservation of $L^2$-norm of $u_{\tau}$ with 	\eqref{apriori-estimate-H1L2-proof-b} and \eqref{apriori-estimate-H1L2-proof-c} we obtain that 
$$\sup_{0< \tau < 1}\big(\|{u_{\tau}}\|_{L^{\infty}_tH^1_x}+\|v_{\tau}\|_{L^{\infty}_tL^2_x}\big)\le \phi(r),$$
where $\phi(\cdot)$ is a polinomial function with $\phi(0)=0$. This concludes the proof. 
}

We can weaken the hypothesis about the data $v_{0\tau}$ to a certain order of growth and retain the same growth rate for $v_{\tau}$, without convergence being affected in 
Theorem \ref{Convergence-L2-A}, as described in the next result.

\begin{lemma}\label{apriori-estimate-H1L2-Neg-energy}
Let $\big\{({u_0}_{\tau}, {v_0}_{\tau})\big\}_{0< \tau <1}$ be a family of data in the space $H^1\times H^1$  under the  following assump\-tions:
\begin{enumerate}
\item[(i)]  $\displaystyle \sup_{0<\tau <1}E_{\tau}(0)\le 0$, \medskip

\item[(ii)] ${\displaystyle \sup_{0<\tau <1}\|{u_0}_{\tau}\|_{L^2}<\infty}\quad\text{and}\quad \|{v_0}_{\tau}\|_{L^2}=O(\tau^{-1+\gamma}),\; 0< \gamma \ll 1.$
\end{enumerate}
Then we have
\begin{equation}
\sup_{0< \tau <1}\|u_{\tau}\|_{L^{\infty}_tH^1_x}< \infty\quad\text{and}\quad \|v_{\tau}\|_{L^{\infty}_tL^2_x}=O(\tau^{-1+\gamma}),\; 0< \gamma \ll 1.
\end{equation}
\end{lemma}
\dem{Once again we can combine the Gagliardo-Nirenberg inequality, the conservation law \eqref{Conservation Law} and the assumption (i) to obtain 
\begin{equation*}\label{apriori-estimate-H1L2-Neg-energy-proof-a}
	\begin{split}
		\|\p_x u_{\tau}(\cdot, t)\|_{L^2}^2+ \|v_{\tau}(\cdot, t)\|_{L^2}^2 & \le 2\|v_{\tau}(\cdot, t)\|_{L^2}\|u_{\tau}(\cdot, t)\|_{L^4}^2 + E_{\tau}(0)\\
		&\le \|v_{\tau}(\cdot, t)\|^2_{L^2} + \|u_{\tau}(\cdot, t)\|_{L^4}^4\\
		&\le \|v_{\tau}(\cdot, t)\|^2_{L^2} +\le C_{gn}^4\|{u_0}_{\tau}\|_{L^2}^3 \|\p_x u_{\tau}(\cdot, t)\|_{L^2}\\
		&\le \|v_{\tau}(\cdot, t)\|^2_{L^2} +\frac{1}{2}C_{gn}^8\|{u_0}_{\tau}\|_{L^2}^6 + \frac{1}{2}\|\p_x u_{\tau}(\cdot, t)\|_{L^2}^2,
	\end{split}
\end{equation*}	
so
\begin{equation*}\label{apriori-estimate-H1L2-Neg-energy-proof-c}
	\|\p_x u_{\tau}(\cdot, t)\|_{L^2} \le C_{gn}^4\|{u_0}_{\tau}\|_{L^2}^3 \le  C_{gn}^4r^3,
\end{equation*}
because of (ii), and $r:={\displaystyle \sup_{0<\tau <1}\|{u_0}_{\tau}\|_{L^2}}$.

Therefore
\begin{equation}\label{apriori-estimate-H1L2-Neg-energy-proof-d}
\sup_{0< \tau <1}\|u_{\tau}\|_{L^{\infty}_tH^1_x}\lesssim r +r^3. 
\end{equation}	
On the other hand, 
\begin{equation*}\label{apriori-estimate-H1L2-Neg-energy-proof-V-a}
\begin{split}
\|v_{\tau}(\cdot, t)\|_{L^2}&\le D_{\tau}(t)\|{v_0}_{\tau}\|_{L^2}+\frac{1}{\tau}\int_0^tD_{\tau}(t-t')\|u_{\tau}(\cdot, t')\|^2_{L^4}dt'\\
&\le \|{v_0}_{\tau}\|_{L^2} + C_{gn}^2\|{u_0}_{\tau}\|_{L^2}^{3/2}\|\p_xu_{\tau}\|_{L_{[0, t]}^{\infty}L^2_x}^{1/2}\frac{1}{\tau}\int_0^tD_{\tau}(t-t')dt'\\
&\lesssim  \|{v_0}_{\tau}\|_{L^2} + r^3(1-D_{\tau}(t))\\
&\lesssim \|{v_0}_{\tau}\|_{L^2} + r^3,
\end{split}
\end{equation*}
for all $t\ge0$. Then, from (ii) we  conclude that
\begin{equation}\label{apriori-estimate-H1L2-Neg-energy-proof-V-b}
\sup_{0< \tau <1}\|v_{\tau}\|_{L^{\infty}_tL^2_x} = O(\tau^{-1+\gamma}).
\end{equation}	
The estimates \eqref{apriori-estimate-H1L2-Neg-energy-proof-d} and \eqref{apriori-estimate-H1L2-Neg-energy-proof-V-b} give the result.
}

\subsection{Further estimates}
\begin{lemma}\label{lemma-fe-1} Suppose that we have the hypotheses in Lemma \ref{apriori-estimate-H1L2} or those given in Lemma \ref{apriori-estimate-H1L2-Neg-energy} and let $u$ be the global  $H^1$-solution of \eqref{CNLS}, with $\lambda =-1$, for some data $u_0$. Then
\begin{equation}
\big \Vert  D_{\tau}(t)(v_{0\tau}+\vert u\vert^2)\big \Vert_{L^1_TL^2_x} =O(\tau^{\gamma}),\; 0< \gamma \ll 1,\quad\text{for all}\quad 0<\tau<1\quad\text{and}\quad T>0.
\end{equation}
\end{lemma}
\dem{From lemmas \ref{apriori-estimate-H1L2} and \ref{apriori-estimate-H1L2-Neg-energy} we have 
$\displaystyle \sup_{0< \tau <1}\|u_{\tau}\|_{L^{\infty}_tH^1_x}< \infty$, so
\begin{equation*}
\begin{split}
 \big \Vert D_{\tau}(t)(v_{0\tau}+\vert u\vert^2) \big\Vert_{L_T^1 L_x^2}&\le\big \Vert D_{\tau}(t)(\Vert {v_0}_{\tau}\Vert_{L^2_x}+\Vert u\Vert_{L^4_x}^2) \big \Vert_{L^1_T}\\
 &\le \big \Vert D_{\tau}(t)(\Vert {v_0}_{\tau}\Vert_{L^2_x}+\Vert u\Vert_{H^1_x}^2) \big \Vert_{L^1_T}\\
 &\le (\Vert {v_0}_{\tau}\Vert_{L^2_x}+\Vert u\Vert_{L^\infty_tH^1_x}^2)\int_0^Te^{-t/\tau}dt'\\
 &=\tau (1-e^{-T/\tau})(\Vert {v_0}_{\tau}\Vert_{L^2}+\Vert u\Vert_{L^\infty_tH^1_x}^2)\\
 &\lesssim \tau^{\gamma},
\end{split}
\end{equation*}
even though $\Vert {v_0}_{\tau}\Vert_{L^2}= O(\tau^{-1+\gamma}).$
}

The following lemma will allow us to control the growth of the difference between the quadratic wave amplitudes.

\begin{lemma}\label{lemma-fe-2}
Let $T>0$ be given and $u$ the global  $H^1$-solution of \eqref{CNLS} for some data $u_0$. For all $\epsilon>0$ there exists a positive number $\delta=\delta(\epsilon,T)$ such that 
\begin{equation}\label{lemma-fe-2-I}
\frac{1}{\tau}\int_{\boldsymbol{t_{\delta}}}^{t}D_{\tau}(t-s)\Vert u(s)-u(t)\Vert_{L^2_x} ds < \epsilon
\end{equation}
for all $t\in [0, T]$, where $\boldsymbol{t_{\delta}}:=\max\{0,t-\delta\}$. Furthermore, 
\begin{equation}\label{lemma-fe-2-II}
\Big\Vert \frac{1}{\tau} \int_0^{t}D_{\tau}(t-s)\Vert u(s)-u(t)\Vert_{L^2_x} ds \Big\Vert_{L^{\infty}_T}\le
2e^{-\delta/\tau}\Vert u_0\Vert_{L^2}+\epsilon 
\end{equation}
and 
\begin{equation}\label{lemma-fe-2-III}
\Big\Vert \frac{1}{\tau} \int_0^{t}D_{\tau}(t-s)\Vert u(s)-u(t)\Vert_{L^2_x} ds \Big\Vert_{L^1_T}\le
2Te^{-\delta/\tau}\Vert u_0\Vert_{L^2}+\epsilon T.
\end{equation}
\end{lemma}

\dem{Let $\epsilon>0$ be given. We have $u\in C\big([0,T]; L^2\big)$, then $u(\cdot, t)$ is uniformly continuous on the time interval $[0, T]$. So, there is a positive number 
$\delta=\delta(\epsilon,T)$ such that
\begin{equation*}
\Vert u(s_1)-u(s_2)\Vert_{L^2_x}< \epsilon 
\end{equation*}
whenever $\vert s_1-s_2\vert<\delta$. Hence
\begin{equation}\label{lemma-fe-2-proof-a}
\frac{1}{\tau}\int_{\boldsymbol{t_{\delta}}}^{t}D_{\tau}(t-s)\Vert u(s)-u(t)\Vert_{L^2_x} ds <  \frac{\epsilon}{\tau}\int_{\boldsymbol{t_{\delta}}}^{t}D_{\tau}(t-s)ds<\epsilon
\end{equation}
and \eqref{lemma-fe-2-I} is proved. To obtain \eqref{lemma-fe-2-II} we split the integration as follows:
\begin{equation*}\begin{split}
\mathcal{I}(\tau, t)&:=\frac{1}{\tau}\int_0^{t}D_{\tau}(t-s)\Vert u(s)-u(t)\Vert_{L^2_x} ds\\
&=\frac{1}{\tau}\int_0^{\boldsymbol{t_{\delta}}}D_{\tau}(t-s)\Vert u(s)-u(t')\Vert_{L^2_x} ds + \frac{1}{\tau}\int_{\boldsymbol{t_{\delta}}}^{t}D_{\tau}(t-s)\Vert u(s)-u(t)\Vert_{L^2_x} ds, 
\end{split}\end{equation*}
then using \eqref{lemma-fe-2-proof-a} and \eqref{Conservation Law}  we get
\begin{equation*}
\mathcal{I}(\tau, t)<
\begin{cases}
\epsilon,&\text{if}\;\;\;t_{\delta}=0,\medskip \\ 
2\Vert u_0\Vert_{L^2_x}\big( e^{-\delta/\tau} - e^{-t/\tau}\big )+\epsilon, &\text{if}\;\;\;t_{\delta}=t-\delta>0.
\end{cases}
\end{equation*}
Therefore, we conclude that 
$$\mathcal{I}(\tau, t)<2\Vert u_0\Vert_{L^2_x}e^{-\delta/\tau} +\epsilon,\; \text{for all}\;  t\in [0,T],$$
which yields \eqref{lemma-fe-2-II} and consequently
\begin{equation*}
\Big\Vert \frac{1}{\tau}\int_0^{t}D_{\tau}(t-s)\Vert u(s)-u(t)\Vert_{L^2_x} ds \Big\Vert_{L^1_T}
=\int_0^T\mathcal{I}(\tau, t)dt\\
<2Te^{-\delta/\tau}\Vert u_0\Vert_{L^2}+\epsilon T,
\end{equation*}
as claimed in \eqref{lemma-fe-2-III}.
}

\begin{lemma}\label{lemma-fe-3} 
Let $T>0$ be given and $u$ the global  $H^1$-solution of \eqref{CNLS} for some data $u_0$. For all $\epsilon>0$ there exists a positive number 
$\tau_{\epsilon}$ such that 
$$\big\|D_{\tau}(t) \big(|u(t)|^2-|u_0|^2 \big)\big\|_{L^{\infty}_TL^2_x}\lesssim \epsilon,$$
for all $0<\tau<\tau_{\epsilon}.$
\end{lemma}
\dem{Since  $u\in C\big([0, T];\,H^1\big)$  there is a positive number $\delta_{\epsilon}$, depending only on $\epsilon$, such that 
$$\big\||u(t)|^2-|u_0|^2\big\|_{L^2_x}\lesssim \epsilon \quad \text{for all}\quad 0\le t \le \delta_{\epsilon}$$
and consequently
\begin{equation}\label{lemma-fe-3-proof-1}
\sup\limits_{0\le t\le \delta_{\epsilon}}\big\|D_{\tau}(t) \big(|u(t)|^2-|u_0|^2 \big)\big\|_{L^2_x}\lesssim \epsilon.
\end{equation}

On the other hand, on the interval $[\delta_{\epsilon}, T]$, using the  Gagliardo-Nirenberg inequality an the conservation law \eqref{Conservation Law} for the cubic NLS equation, 
we have 
\begin{equation*}
\begin{split}
\big\|D_{\tau}(t)|u(t)|^2-|u_0|^2\big\|_{L^2_x}&=e^{-t/\tau}\big\||u(t)|^2-|u_0|^2\big\|_{L^2_x}\\
&\le e^{-\delta_{\epsilon}/\tau}c(\|u_0\|_{H^1}),
\end{split}
\end{equation*}
for all $\delta_{\epsilon}\le t \le T$, where $c$ is constant depending only on $\|u_0\|_{H^1}$. Hence, we conclude that there is $\tau_{\epsilon}>0$ such that 
\begin{equation}\label{lemma-fe-3-proof-2}
\sup\limits_{\delta_{\epsilon}\le t\le T}\big\|D_{\tau}(t) \big(|u(t)|^2-|u_0|^2 \big)\big\|_{L^2_x}\lesssim \epsilon,
\end{equation}
for all $0<\tau<\tau_{\epsilon}$.

The proof of the claimed  result follows directly from the estimates \eqref{lemma-fe-3-proof-1} and \eqref{lemma-fe-3-proof-2}. 
}

\section{\bf{Local theory in $\boldsymbol{L^2(\R)\times L^p(\R),\; 1\le p < \infty}$.}}

Here we show  the proof of well-posedness results for data in $L^2(\R)\times L^p(\R)$. The technique used to obtain them is the classical 
fixed point procedure combined with the use of Strichartz estimates for the unitary group $S(t)=e^{it\partial_x^2/2}$ in the one spatial dimension. 
See, for example, the books \cite{Cazenave-Book, Linares-Ponce} for the use of similar technique in the context of non-linear Schr\"odinger equation.  

Technically, we divided the proof into two cases: $1\le p<2$  and $2\le p <\infty$. The reason is that the functional space, where the solutions will be obtained, 
is defined by using different mixed $L^r_TL^q_x$-spaces, admissible for Strichartz estimates depending of each one of the cases. 

\subsection{\bf{Proof of Theorem \ref{LWP-SD}}}
Consider de integral formulation  for (\ref{SD}), given by
\begin{equation*}
\begin{cases}
\displaystyle u(\cdot, t) = S(t)u_0-i\int_0^tS(t-s)u(\cdot, s)v(\cdot, s)ds,\medskip\\
\displaystyle v(\cdot, t) = e^{-t/\tau}v_0+\tfrac{\lambda}{\tau}\int_0^t e^{-\frac{t-s}{\tau}}|u(\cdot, s)|^2ds,
\end{cases}
\end{equation*}
from which we define the two operators
\begin{align}
&\Phi_1(u,v):=S(t)u_0-i\int_0^tS(t-s)u(\cdot, s)v(\cdot, s)ds,\label{Sd-Integral-u}\\
&\Phi_2(u,v):=e^{-t/\mu}v_0+\tfrac{\lambda}{\tau}\int_0^t e^{-\frac{t-s}{\tau}}|u(\cdot, s)|^2ds\label{Sd-Integral-v}.
\end{align}

We divide the proof into two cases. 

\medskip
\noindent{\bf{\underline{Case:}} $\boldsymbol{1\le p< 2}.$} Let $a$ and $b$ positive numbers that will be determined later.  Consider the sets 
\begin{align}
&U_{a,T}=\bigg\{u: [0,T]\times \R\rightarrow \C;\; \|u\|_U:=\|u\|_{L_T^{\infty}L^2_x} + \|u\|_{L^{\frac{4p}{2-p}}_TL^{\frac{p}{p-1}}_x} + \|u\|_{L^{\frac{4p}{p-1}}_TL^{2p}_x}\le a\bigg\}\\
\intertext{and}
&V_{b,T}=\Big\{v: [0,T]\times \R \rightarrow \R;\; \|v\|_V:=\|v\|_{L_T^{\infty}L^p_x}\le b\Big\}.
\end{align}
Note that in the case $p=1$ we have $\|u\|_{L^{\frac{4p}{p-1}}_TL^{2p}_x}=\|u\|_{L_T^{\infty}L^2_x}.$

As usual, we will next choose $a$, $b$ and $T$ so that the operator $\Phi=(\Phi_1, \Phi_2)$ maps
$U_{a,T}\times V_{b,T}$ to itself, 
$$\Phi=(\Phi_1, \Phi_2):U_{a,T}\times V_{b,T} \longrightarrow  U_{a,T}\times V_{b,T},$$
and it is a contraction with the norm
\begin{equation}\label{proof-lwp-1d}
\|(u,v)\|_{U\times V}=\|u\|_U + \|v\|_V,
\end{equation}
yielding the fixed point that satisfies the integral formulation of the problem. 

Indeed, note that
\begin{equation*}\label{proof-lwp-I-a}
\begin{split}
\|\Phi_1(u,v)\|_{U}&\le \|S(t)u_0\|_U + \Bigl \|\int_0^tS(t-s)u(\cdot, s)v(\cdot, s)ds \Bigl\|_U\\
&\le c\|u_0\|_{L^2} + c\|uv\|_{L^{4/3}_TL^1_x}.
\end{split}
\end{equation*}
This follows, for the homogeneous term, by (\ref{Strichart-homogeneous}) with $(p_1,q_1)=(\infty, 2)$, $(p_1,q_1)=(\frac{4p}{2-p},\frac{p}{p-1})$ and $(p_1,q_1)=(\frac{4p}{p-1},2p)$ with $1< p<2$. For the non-homogeneous
term we use (\ref{Strichart-non-homogeneous}) with the same pairs $(p_1,q_1)$ chosen in the previous case and with $(p_2,q_2)=(4,\infty)$.

Now, using  H\"older's inequality we obtain, for all $(u,v)\in U_{a,T}\times V_{b,T}$, the following estimates:
\begin{equation}\label{proof-lwp-I-b}
\begin{split}
\|\Phi_1(u,v)\|_{U}&\le c \|u_0\|_{L^2} + c\|u\|_{L^{\frac{4p}{2-p}}_TL^{\frac{p}{p-1}}_x}\|v\|_{L^{\frac{2p}{2p-1}}_TL^p_x}\\
& \le c \|u_0\|_{L^2} + cT^{\frac{2p-1}{2p}}\|u\|_{L^{\frac{4p}{2-p}}_TL^{\frac{p}{p-1}}_x}\|v\|_{L^{\infty}_TL^p_x}\\
&\le c \|u_0\|_{L^2} + cT^{\frac{2p-1}{2p}}ab.
\end{split}
\end{equation}

On the other hand, for $1\le p <2$, applying the Minkowski's inequality to \eqref{Sd-Integral-v} we get
\begin{equation*}\label{proof-lwp-I-c}
\|\Phi_2(u,v)\|_{L^p_x}\le e^{-t/\tau}\|v_0\|_{L^p} + \tfrac{1}{\tau}\int_0^t e^{-\frac{t-s}{\tau}}\| |u|^2 \|_{L^p_x}ds;\\
\end{equation*}
then  using the exponential decay of the free propagator of the Debye and  H\"older's inequality, its follows  that
\begin{equation}\label{proof-lwp-I-d}
\begin{split}
\|\Phi_2(u,v)\|_{L^p_x}&\le e^{-t/\tau}\|v_0\|_{L^p} + \tfrac{1}{\tau}\int_0^t\|u\|^2_{L^{2p}_x}ds\\
&\le \|v_0\|_{L^p} + \tfrac{1}{\tau}t^{\frac{p+1}{2p}}\left(\int_0^t\|u\|^{\frac{4p}{p-1}}_{L^{2p}_x}ds \right)^{\frac{p-1}{2p}}.
\end{split}
\end{equation} 
Hence, we conclude that 
\begin{equation}\label{proof-lwp-I-e}
\begin{split}
\|\Phi_2(u,v)\|_{L^{\infty}_TL^p_x}&\le \|v_0\|_{L^p} + \tfrac{T^{\frac{p+1}{2p}}}{\tau}\|u\|^2_{L_T^{\frac{4p}{p-1}}L_x^{2p}}\\
&\le  \|v_0\|_{L^p}+ \tfrac{T^{\frac{p+1}{2p}}}{\tau}a^2.
\end{split}
\end{equation} 
Then, taking $a:=2c\|u_0\|_{L^2}$ and  $b:=2\|v_0\|_{L^p}$ and combining the inequalities \eqref{proof-lwp-I-b} and \eqref{proof-lwp-I-e} we obtain 
$$\Phi_1(u,v)\le a\quad \text{and}\quad \Phi_2(u,v)\le b,$$
for all
$$T\le T_1:=\min\left\{\left(\frac{1}{4c\|v_0\|_{L^p}}\right)^{\frac{2p}{2p-1}},\;\left(\frac{\tau\|v_0\|_{L^p}}{4c^2\|u_0\|^2_{L^2}}\right)^{\frac{2p}{p+1}}\right\}.$$
Thus, $(\Phi_1(u,v),\;\Phi_2(u,v)) \in U_{a,T}\times V_{b,T}$. On the other hand, similar arguments yield  the following contraction estimates:
\begin{equation*}
\begin{split}
\|\Phi_1(u_1,v_1)-\Phi_1(u_2,v_2)\|_U& \le  cT^{\frac{2p-1}{2p}}\left(\|u_1-u_2\|_{L^{\frac{4p}{2-p}}_TL^{\frac{p}{p-1}}_x}\|v_1\|_{L^{\infty}_TL^p_x}+ \|u_2\|_{L^{\frac{4p}{2-p}}_TL^{\frac{p}{p-1}}_x}\|v_1-v_2\|_{L^{\infty}_TL^p_x}\right)\\
&\le  cT^{\frac{2p-1}{2p}}(a+b)\|(u_1,v_1)-(u_2,v_2)\|_{U\times V}
\end{split}
\end{equation*}
and 
\begin{equation*}
\begin{split}
\|\Phi_2(u_1,v_1)-\Phi_2(u_2,v_2)\|_V & \le \tfrac{T^{\frac{p+1}{2p}}}{\tau}\|u_1-u_2\|_{L_T^{\frac{4p}{p-1}}L_x^{2p}}\|u_1+u_2\|_{L_T^{\frac{4p}{p-1}}L_x^{2p}}\\
&\le 2a\tfrac{T^{\frac{p+1}{2p}}}{\tau}\|(u_1,v_1)-(u_2,v_2)\|_{U\times V}.\\
\end{split}
\end{equation*}
Hence,
\begin{equation*}
\begin{split}
\|\Phi(u_1, v_1) - \Phi(u_2, v_2)\|_{U\times V}& =\|\Phi_1(u_1,v_1)-\Phi_1(u_2,v_2)\|_U+\|\Phi_2(u_1,v_1)-\Phi_2(u_2,v_2)\|_V\\
&\le \frac12 \|(u_1, v_1)-(u_2, v_2)\|_{U\times V}
\end{split}
\end{equation*}
for all 
$$T\le T_2:=\min\left\{\left(\frac{1}{8c(c\|u_0\|_{L^2}+\|v_0\|_{L^p})}\right)^{\frac{2p}{2p-1}},\;\left(\frac{\tau}{16c\|u_0\|_{L^2}}\right)^{\frac{2p}{p+1}} \right\}.$$
Thus, we can conclude that the operator $\Phi$ has a unique fixed point in the set $U_{a,T}\times V_{b,T}$ with
$$T(\tau, \|u_0\|_{L^2}, \|v_0\|_{L^p}):= \min\big\{ T_1, T_2 \big\}.$$
The rest of conclusions of the theorem follow from the standard arguments, as in the non-linear Schr\"odinger equation.   

\medskip
\noindent{\bf{\underline{Case:}} $\boldsymbol{2\le p< \infty}.$} The situation in this case is very similar to the previous one and we only show the sketch  of the proof. 

We proceed as in the previous case with an slight  change in the definition of $U_{a, T}$, more precisely:
$$
U_{a,T}=
\bigg\{u: [0,T]\times \R\rightarrow \C;\; \|u\|_U:=\|u\|_{L_T^{\infty}L^2_x} + \|u\|_ {L^{2p}_TL^{\frac{2p}{p-2}}_x} + \|u\|_{L^{\frac{4p}{p-1}}_TL^{2p}_x}\le a\bigg\} 
$$
Using (\ref{Strichart-homogeneous}) with $(p_1,q_1)=(\infty, 2)$,  $(p_1,q_1)=\big(\frac{4p}{p-1},2p\big)$ and $(p_1,q_1)=\big(2p,\frac{2p}{p-2}\big)$ with $2\le  p< \infty$ and  (\ref{Strichart-non-homogeneous}) with the same pairs $(p_1,q_1)$ and with $(p_2,q_2)=(\infty, 2)$, we get 
\begin{equation*}\label{proof-lwp-II-a}
\begin{split}
\|\Phi_1(u,v)\|_{U}&\le \|S(t)u_0\|_U + \Bigl \|\int_0^tS(t-s)u(\cdot, s)v(\cdot, s)ds \Bigl\|_U\\
&\le c\|u_0\|_{L^2} + c\|uv\|_{L^1_TL^2_x}.
\end{split}
\end{equation*}
Then, applying  H\"older's inequality we obtain
\begin{equation*}\label{proof-lwp-II-b}
\begin{split}
\|\Phi_1(u,v)\|_{U}&\le c \|u_0\|_{L^2} + c\|u\|_{L^{2p}_TL^{\frac{2p}{p-2}}_x}\|v\|_{L^{\frac{2p}{2p-1}}_TL^p_x}\\
& \le c \|u_0\|_{L^2} + cT^{\frac{2p-1}{2p}}\|u\|_{L^{2p}_TL^{\frac{2p}{p-2}}_x}\|v\|_{L^{\infty}_TL^p_x}\\
&\le c \|u_0\|_{L^2} + cT^{\frac{2p-1}{2p}}ab.
\end{split}
\end{equation*}

The estimate for $\|\Phi_2(u,v)\|_{L^{\infty}_TL^p_x}$ remains the same as \eqref{proof-lwp-I-e}. The rest of the proof proceeds in the same way as the previous case.

\subsection{Proof of Corollary \ref{GWP-SD}} From \eqref{Conservation Law} we have the control of $L^2$ norm of $u$ and also, from \eqref{equation-v}, we have that
\begin{equation*}
\begin{split}
\|v(\cdot, t)\|_{L^1} &\le \|e^{-t/\tau}v_0\|_{L^1}+\frac{1}{\tau}\int_0^te^{-\frac{t-s}{\tau}}\big\| |u(\cdot, s)|^2 \big\|_{L^1}ds\\
&\le e^{-t/\tau}\|v_0\|_{L^1} + (1-e^{-t/\tau})\|u_0\|^2_{L^2}\\
&\le \|v_0\|_{L^1} + \|u_0\|^2_{L^2}.
\end{split}
\end{equation*}
Then,  by standard arguments this a priori  bound  for the $L^2\times L^1$-norm  ensures that the local solution can be extended to any time interval $[0, T]$. 

\section{\textbf{Proof of the convergence results}}

\subsection{Proof of Theorem \ref{Convergence-L2-A}}
In order to simplify the explanation we use the notation $u_{\tau}(t):=u_{\tau}(\cdot, t)$ and the same for $v_{\tau}$ and $u$. 

Let $T^*$ a fixed positive time and let $\delta>0$ be as in the Lemma \ref{lemma-fe-2} concerning to the interval $[0, T^*]$, which is associated with a positive $\epsilon>0$ given. Also consider a very small number $\gamma$ ($0< \gamma \ll 1$).

Next we shall estimate the difference between the solutions $u_{\tau}$ and $u$ using the Duhamel formulation, that is
\begin{equation}\label{Convergence-L2-A-proof-1}
u_{\tau}(t)-u(t)=S(t)(u_{0\tau}-u_0)-i\int_0^{t}S(t-s)\big(u_{\tau}(s)v_{\tau}(s)-\lambda\vert u(s)\vert^2 u(s)\big)ds,
\end{equation}
where $0\le t\le T^*$.   

We will be performed the next computations on the interval $[0, T]$ with $T\in (0, T^*)$  which will be chosen later. The nonlinear integral term can be writed as
\begin{equation}\label{Convergence-L2-A-proof-2}
\begin{split}
\mathcal{N}(\tau; \cdot, t)&:=-i\int_0^{t}S(t-s)\big[u_{\tau}(s)v_{\tau}(s)-\lambda\vert u(s)\vert^2 u(s)\big]ds\\
&=-i\int_0^{t}S(t-s)u_{\tau}(s)\big[v_{\tau}(s)-\lambda\vert u(s)\vert^2\big]ds-i\int_0^{t}S(t-s)\big[u_{\tau}(s)-u(s)\big]\lambda\vert u(s)\vert^2ds\\
&:=\mathcal{N}_1(\tau; \cdot, t)+\mathcal{N}_2(\tau; \cdot, t).
\end{split}
\end{equation}

In what follows we consider $\lambda =-1$. In any situations of the hypotheses of Theorem \ref{Convergence-L2-A}, the lemmas \ref{apriori-estimate-H1L2} and  \ref{apriori-estimate-H1L2-Neg-energy} ensure that
\begin{equation}\label{Convergence-L2-A-proof-2a} 
a_0:=\sup\limits_{0<\tau<1}\Vert u_{\tau}\Vert_{L^\infty_tH^1_x}<\infty\quad \text{and}\quad 
\sup\limits_{0< \tau <1}\|v_{\tau}\|_{L^{\infty}_tL^2_x} = O(\tau^{-1+\gamma}),\; 0<\gamma \ll 1.
\end{equation}
Then, by Sobolev embedding we have 
\begin{equation}\label{Convergence-L2-A-proof-3}
\begin{split}
\Vert \mathcal{N}_1(\tau; \cdot, t)\Vert_{L_T^\infty L_x^2}&\le \big\Vert u_{\tau}\big(v_{\tau}+|u|^2\big)\big\Vert_{L_T^1 L_x^2}\\
&\le \Vert u_{\tau}\Vert_{L_t^{\infty}L^{\infty}_x}\Vert v_{\tau}+|u|^2\Vert_{L_T^1 L_x^2}\\
&\le a_0\big\Vert v_{\tau}+|u|^2\big\Vert_{L_T^1 L_x^2}.
\end{split}
\end{equation}
Now we need to estimate $\big\Vert v_{\tau}+\vert u\vert^2\big\Vert_{L_T^1 L_x^2}$ and to do this we first observe that 
\begin{equation}\label{Convergence-L2-A-proof-4}
v_{\tau}(t)+ |u(t)|^2=
D_{\tau}(t)\big({v_0}_{\tau}+|u(t)|^2\big)-\frac{1}{\tau}\int_0^tD_{\tau}(t-s)\big[|u_{\tau}(s)|^2 -|u(t)|^2\big]ds;
\end{equation}
then using this equality and Lemma \ref{lemma-fe-1} we get
\begin{equation}\label{Convergence-L2-A-proof-5}
\begin{split}
\big\Vert v_{\tau}+\vert u\vert^2\big\Vert_{L_T^1 L_x^2}
&\le \big\Vert D_{\tau}(t)\big({v_0}_{\tau}+\vert u\vert^2\big)\big\Vert_{L_T^1L_x^2}
+\frac{1}{\tau}\Big\Vert\int_0^{t}D_{\tau}(t-s)\big[\vert u_{\tau}(s)\vert^2-\vert u(t)\vert^2\big]ds\Big\Vert_{L_T^1 L_x^2}\\
&\le O(\tau^{\gamma})
+\frac{1}{\tau}\Big\Vert\int_0^{t}D_{\tau}(t-s)\big\|\vert u_{\tau}(s)\vert^2-\vert u(t)\vert^2\big\|_{L_x^2}ds\Big\Vert_{L_T^1}\\
&= O(\tau^{\gamma}) + \mathcal{N}_3(\tau, T).
\end{split}
\end{equation}
Now using the Sobolev embedding we have
\begin{equation}\label{Convergence-L2-A-proof-6}
\begin{split}
\big\|\vert u_{\tau}(s)\vert^2-\vert u(t)\vert^2\big\|_{L_x^2}
 &\le \big\Vert \vert u_{\tau}(s)\vert-\vert u(t)\vert \big\Vert_{L^2_x} \Vert \vert u_{\tau}(s)\vert+\vert u(t)\vert\big\Vert_{L^\infty_x}\\
 &\le \big\Vert u_{\tau}(s)- u(t)\big\Vert_{L^2_x}\big(\Vert u_{\tau}\Vert_{L^\infty_tH^1_x}+\Vert u\Vert_{L^\infty_tH^1_x}\big)\\
 &\le b_0\big(\Vert u_{\tau}(s)- u(s)\Vert_{L^2_x}+ \Vert u(s)- u(t)\Vert_{L^2_x}\big),
\end{split}
\end{equation}
where 
$$b_0:=a_0 + \Vert u\Vert_{L^\infty_tH^1_x}\le a_0 +  c\|u_0\|_{H^1}:= c_0.$$
Thus, from \eqref{Convergence-L2-A-proof-6} and Lemma \ref{lemma-fe-2}
it follows that
\begin{equation}\label{Convergence-L2-A-proof-7}
\begin{split}
 \mathcal{N}_3(\tau, T)\le& \frac{b_0}{\tau}\Big\Vert \int_0^{t}D_{\tau}(t-s)\Vert u_{\tau}(s)- u(s)\Vert_{L^2_x}ds \Big\Vert_{L^1_T}+\frac{b_0}{\tau}\Big\Vert \int_0^{t}D_{\tau}(t-s)\Vert u(s)- u(t)\Vert_{L^2_x}ds \Big\Vert_{L^1_T}\\
 &\le b_0T\big(\Vert u_{\tau}-u\Vert_{L^{\infty}_TL^2_x}+2e^{-\delta/\tau}\Vert u_0\Vert_{L^2}+\epsilon \big).
 \end{split}
\end{equation}
Collecting the informations in \eqref{Convergence-L2-A-proof-3}, \eqref{Convergence-L2-A-proof-5} and \eqref{Convergence-L2-A-proof-7} we conclude that
\begin{equation}\label{Convergence-L2-A-proof-8}
\begin{split}
\Vert \mathcal{N}_1(\tau;\cdot, t)\Vert_{L_T^\infty L_x^2}&\le O(\tau^{\gamma}) + a_0b_0T\big(\Vert u_{\tau}-u\Vert_{L^{\infty}_TL^2_x}
+2e^{-\delta/\tau}\Vert u_0\Vert_{L^2}+\epsilon \big).
\end{split}
\end{equation}
Using the Sobolev embedding we can estimate the term $\mathcal{N}_2(\tau, t)$ as follows:
\begin{equation}\label{Convergence-L2-A-proof-9}
\begin{split}
\Vert \mathcal{N}_2(\tau; \cdot, t)\Vert_{L^{\infty}_TL_x^2}&\le \Big\|\int_0^{t}\|u_{\tau}(s)-u(s)\|_{L^2_x}\|u(s)\|^2_{L^{\infty}_x}ds\Big\|_{L^{\infty}_T} \\
&\le T \Vert u\Vert^2_{L^{\infty}_tH^1_x}\Vert u_\tau -u\Vert_{L^{\infty}_TL^2_x}.\\
&\le c_0^2 T \Vert u_\tau -u\Vert_{L^{\infty}_TL^2_x}.
\end{split}
\end{equation}

From  \eqref{Convergence-L2-A-proof-1}, \eqref{Convergence-L2-A-proof-8} and \eqref{Convergence-L2-A-proof-9} we obtain the estimate
\begin{equation*}
\Vert u_{\tau}-u\Vert_{L^{\infty}_TL^2_x}(1-d_0T)\lesssim \Vert {u_0}_{\tau}-u_0\Vert_{L^2_x}+O(\tau^{\gamma})+2Te^{-\delta/\tau}\Vert u_0\Vert_{L^2}+\epsilon T,
\end{equation*}
where $d_0$ and the others unspecified constants, depend only on the norm of $\|u_0\|_{H^1}$ and on the radius of the ball where initial data ${u_0}_{\tau}$ have been chosen. 

Now we set $T=\dfrac{1}{2d_0}$ and then, choosing $\tau$ small enough, we obtain
\begin{equation}
\Vert u_{\tau}-u\Vert_{L^{\infty}_TL^2_x}\lesssim\epsilon.
\end{equation} 

Notice that we can repeat the same arguments as before in the interval $[T,2T]$ to obtain
\begin{equation}
\Vert u_{\tau}-u\Vert_{L^{\infty}_{2T}L^2_x}\le \Vert u_{\tau}-u\Vert_{L^{\infty}_{[0,T]}L^2_x} + \Vert u_{\tau}-u\Vert_{L^{\infty}_{[T,2T]}L^2_x}\lesssim \epsilon,
\end{equation}
for a $\tau$ small enough. So, after applying this procedure a finite number of times we reach the  interval $[0,T^*]$ with small growth of the $L^2$-norm, more precisely
\begin{equation}
\Vert u_{\tau}-u\Vert_{L^{\infty}_{T^*}L^2_x}\lesssim\epsilon
\end{equation}
for all $\tau$ suitable small.

To finish, we observe that as in \eqref{Convergence-L2-A-proof-5}
\begin{equation*}
\big\Vert v_{\tau}+\vert u\vert^2\big\Vert_{L^1_{T^*}L_x^2}
\lesssim O(\tau^{\gamma})+T^*\Vert u_{\tau}-u\Vert_{L^{\infty}_{T^*}L^2_x} +2T^* e^{-\delta/\tau}\Vert u_0\Vert_{L^2}+\epsilon T^*
\lesssim \epsilon
\end{equation*}
for all $\tau$ small enough. This completes the proof.

\begin{remark}
In the case of negative energy and initial data ${v_0}_{\tau}$ increasing with order $O(\tau^{-1+\gamma})$, we have a slower convergence rate because of the term of the same order. For uniformly bounded data  the convergence rate is a bit faster, because this reaches $O(\tau^{\gamma})$ in the initial layer.
\end{remark}

\subsection{Proof of Theorem \ref{Convergence-L2-B}} \label{section-proof-convergence-L2-B}
The statement in \eqref{Convergence-L2-B-1} has  already been  obtained in the proof of Theorem \ref{Convergence-L2-A}. 
The key point in the proof of \eqref{Convergence-L2-B-2} is the use of the equality:
\begin{equation*}
\begin{split}
\mathcal{G}(\tau; \cdot, t)&:=v_{\tau}(t) +|u(t)|^2-e^{-t/\tau}\big({v_0}_{\tau} + |u_0|^2\big)\\
&=e^{-t/\tau}\big(|u(t)|^2 - |u_0|^2\big) -\frac1{\tau}\int_0^tD_{\tau}(t-s)\big( |u_{\tau}(s)|^2 -|u(t)|^2\big)ds\\
&=e^{-t/\tau}\big(|u(t)|^2 - |u_0|^2\big)\\
&\qquad \quad -\frac1{\tau}\int_0^tD_{\tau}(t-s)\big( |u_{\tau}(s)|^2 -|u(s)|^2\big)ds -\frac1{\tau}\int_0^tD_{\tau}(t-s)\big( |u(s)|^2 -|u(t)|^2\big)ds\\
&:=D_{\tau}(t)\big(|u(t)|^2 - |u_0|^2\big) + \mathcal{G}_1(\tau; \cdot, t) + \mathcal{G}_2(\tau; \cdot, t),
\end{split}
\end{equation*}
with $0\le t\le T^*$.

Now we consider a given positive number $\epsilon$. From Lemma \ref{lemma-fe-3} we have that there exits a positive $\tau_1(\epsilon)$ verifying that 
\begin{equation}\label{Convergence-L2-B-proof-1}
\big\|D_{\tau}(t)\big(|u(t)|^2 - |u_0|^2\big) \big\|_{L^{\infty}_{T^*}L^2_x}\lesssim \epsilon,\quad \text{for all}\quad  0< \tau < \tau_1(\epsilon).
\end{equation}

Now we proceed with the estimation of the integral term $\mathcal{G}_1(\tau; \cdot, t)$.  From Sobolev embedding  and lemmas \ref{apriori-estimate-H1L2} and \ref{apriori-estimate-H1L2-Neg-energy} we have
\begin{equation*}\label{Convergence-L2-B-proof-2}
\begin{split}
\big\|\mathcal{G}_1(\tau; \cdot, t)\big\|_{L^2_x}&\le
\frac1{\tau}\int_0^tD_{\tau}(t-s)\big\| |u_{\tau}(s)| -|u(s)| \big\|_{L^2_x}\big\| |u_{\tau}(s)| +|u(s)| \big\|_{L^{\infty}_x}ds\\
&\le c(\|{u_0}_{\tau}\|_{H^1}, \|{v_0}_{\tau}\|_{L^2}, \|u_0\|_{H^1})\big\| u_{\tau} -u \big\|_{L^{\infty}_{T^*}L^2_x}\,\frac1{\tau}\int_0^tD_{\tau}(t-s)ds\\
&= c(\|{u_0}_{\tau}\|_{H^1}, \|{v_0}_{\tau}\|_{L^2}, \|u_0\|_{H^1})\left( 1- e^{-t/\tau}\right)\big\| u_{\tau} -u \big\|_{L^{\infty}_{T^*}L^2_x}\\
&\lesssim \big\| u_{\tau} -u \big\|_{L^{\infty}_{T^*}L^2_x},
\end{split}
\end{equation*}
for all $0\le t\le T^*$. Then, using the convergence $\big\|u_{\tau} -u\big\|_{L^{\infty}_{T^*}L^2_x} \rightarrow 0$ as $\tau \to 0$, we conclude that there exists 
a positive number $\tau_2(\epsilon)$ such that 
\begin{equation}\label{Convergence-L2-B-proof-3}
\big\|\mathcal{G}_1(\tau; \cdot, t)\big\|_{L^{\infty}_{T^*}L^2_x}\lesssim \epsilon,\quad \text{for all}\quad 0< \tau < \tau_2(\epsilon).
\end{equation}

On the other hand, the estimate \eqref{lemma-fe-2-II} in Lemma \ref{lemma-fe-2} give us that there is $\delta=\delta(\epsilon, T^*)$ such that
\begin{equation*}\label{Convergence-L2-B-proof-4}
\begin{split}
\big\|\mathcal{G}_2(\tau; \cdot, t)\big\|_{L^2_x}&\le
\frac1{\tau}\int_0^tD_{\tau}(t-s)\big\| |u(s)| -|u(t)| \big\|_{L^2_x}\big\| |u(s)| +|u(t)| \big\|_{L^{\infty}_x}ds\\
&\le c(\|u_0\|_{H^1})\Big\Vert \frac{1}{\tau} \int_0^{t}D_{\tau}(t-s)\Vert u(s)-u(t)\Vert_{L^2_x} ds \Big\Vert_{L^{\infty}_{T^*}}\\
&\lesssim 2e^{-\delta/\tau}\Vert u_0\Vert_{L^2}+\epsilon,
\end{split}
\end{equation*}
for all $0\le t\le T^*$. Hence, there exists a positive number $\tau_3(\epsilon)$ such that 
\begin{equation}\label{Convergence-L2-B-proof-5}
\big\|\mathcal{G}_2(\tau; \cdot, t)\big\|_{L^{\infty}_{T^*}L^2_x}\lesssim \epsilon,\quad \text{for all}\quad 0< \tau < \tau_3(\epsilon).
\end{equation}

From \eqref{Convergence-L2-B-proof-1}, \eqref{Convergence-L2-B-proof-3} and \eqref{Convergence-L2-B-proof-5} we have that 
$$\big\|v_{\tau}+|u|^2-e^{-t/\tau}\big({v_0}_{\tau} + |u_0|^2\big)\big\|_{L^{\infty}_{T^*}L^2_x}\lesssim \epsilon,$$
for all $0 < \tau < \tau^*(\epsilon)=\min \big\{\tau_1(\epsilon), \tau_2(\epsilon), \tau_3(\epsilon)\big\}$, which give us \eqref{Convergence-L2-B-2}.

Finally, in the case of compatibility condition we have  
\begin{equation*}\label{Convergence-L2-B-proof-6}
\begin{split}
\big\|v_{\tau} +|u|^2\big\|_{L^{\infty}_{T^*}L^2_x}&\le
\big\|v_{\tau} +|u|^2-e^{-t/\tau}\big({v_0}_{\tau} + |u_0|^2\big)\big\|_{L^{\infty}_{T^*}L^2_x} 
+ \big\|e^{-t/\tau}\big({v_0}_{\tau} + |u_0|^2\big)\big\|_{L^{\infty}_{T^*}L^2_x}\\
&\le \big\|v_{\tau} +|u|^2-e^{-t/\tau}\big({v_0}_{\tau} + |u_0|^2\big)\big\|_{L^{\infty}_{T^*}L^2_x} + \big\|{v_0}_{\tau} + |u_0|^2\big\|_{L^2_x},
\end{split}
\end{equation*}
so $\displaystyle \lim\limits_{\tau \to 0}\big\|v_{\tau} +|u|^2\big\|_{L^{\infty}_{T^*}L^2_x}=0$. 
\section{\textbf{Final remarks}}

\begin{remark}[\textbf{Convergence in $\boldsymbol{L^{\infty}_{T^*}\times L^1_x}$}]
If we suppose that initial data $\big({u_0}_{\tau}, {v_0}_{\tau}\big)$ are taken in the space $\in H^1\times (H^1\cap L^1)$, then the integral equation \eqref{equation-v} guarantees that solutions $v_{\tau}$ remain in $L^1$. In this situation the enunciate of Theorem \ref{Convergence-L2-B} worths in the space $L^{\infty}_{T^*}L^2_x\times  L^{\infty}_{T^*}L^1_x$
instead $L^{\infty}_{T^*}L^2_x\times  L^{\infty}_{T^*}L^2_x$. Indeed, Lemma \ref{lemma-fe-3} can be adapted easily to the norm $L^{\infty}_{T}L^1_x$ as well as the proof of Theorem \ref{Convergence-L2-B} follows analogously to the one shown in Section \ref{section-proof-convergence-L2-B}. 
\end{remark}

\begin{remark}[\textbf{Convergence in the periodic setting}]
	We note that the convergence results in theorems \ref{Convergence-L2-A} and \ref{Convergence-L2-B} are also valid in the periodic case ($x\in \T$) and the proof
	follows in a similar way without significant changes. Also, in view of Theorem B (local theory for SD with periodic initial data), enunciated in the introduction, 
	we only need to consider solutions in $H^1(\T)\times L^2(\T)$ since this regularity os covered by the local theory and consequently, from the pseudo-Hamiltonian structure, these solutions are global in time. 
\end{remark}

\section*{\textbf{Acknowledgments}}
A. Corcho  would like to thank the kind hospitality of Department of Mathematics and Statistics of the National University of Colombia-Manizales,
where part of this work was developed. J. Cordero thanks to the Insitituto de Matemática Pura e Aplicada - IMPA by the sopport during the summer posdoctoral position of 2017, where this work can be finished. Finally, the authors would like to thank to professor F. Linares for his valuable comments on this work.

\end{document}